\documentclass[aps,prx,reprint,amsmath,amssymb,superscriptaddress, nofootinbib, 10pt]{revtex4-2}

\usepackage[usenames,dvipsnames]{xcolor}
\usepackage{amssymb}
\usepackage{graphicx}
\usepackage{amsmath}
\usepackage{braket}
\usepackage[pdfusetitle,bookmarks=true,colorlinks,citecolor=blue,urlcolor=blue]{hyperref}

\graphicspath{{Figures/}}

\newcommand{\be}{\begin{equation}}
\newcommand{\ee}{\end{equation}}

\newcommand{\dw}{\downarrow}
\newcommand{\up}{\uparrow}

\newcommand{\eqw}[1]{(\ref{#1})}

\newcommand{\fig}[1]{Fig.\thinspace{}\ref{#1}}

\newcommand{\fc}[1]{({#1})}
\newcommand{\figc}[2]{Fig.\thinspace{}\ref{#1}\thinspace{}\fc{#2}}

\newcommand{\App}[1]{Appendix \ref{#1}}

\begin{document}

\newcommand{\titleinfo}{
Observation of magnon bound states in the long-range, anisotropic Heisenberg model
}

\title{\titleinfo}

\newcommand{\TUM}{\affiliation{Department of Physics, Technical University of Munich, 85748 Garching, Germany}}
\newcommand{\MCQST}{\affiliation{Munich Center for Quantum Science and Technology (MCQST), Schellingstr. 4, 80799 M{\"u}nchen, Germany}}
\newcommand{\IQOQI}{\affiliation{Institute for Quantum Optics and Quantum Information, Austrian Academy of Sciences, Technikerstra{\ss}e 21a, 6020 Innsbruck, Austria}}
\newcommand{\UIBK}{\affiliation{Institut f\"ur Experimentalphysik, Universit\"at Innsbruck, Technikerstra{\ss}e 25, 6020 Innsbruck, Austria}}
\newcommand{\AQT}{\affiliation{AQT, Technikerstra{\ss}e 17, 6020 Innsbruck, Austria}}

\author{Florian Kranzl}
\email{The first two coauthors contributed equally.}
\IQOQI
\UIBK

\author{Stefan Birnkammer}
\email{The first two coauthors contributed equally.}
\TUM
\MCQST

\author{Manoj K. Joshi}
\IQOQI

\author{Alvise Bastianello}
\TUM
\MCQST

\author{Rainer Blatt}
\IQOQI
\UIBK

\author{Michael Knap}
\TUM
\MCQST

\author{Christian F. Roos}
\IQOQI
\UIBK

\begin{abstract}
Over the recent years coherent, time-periodic modulation has been established as a versatile tool for realizing novel Hamiltonians. Using this approach, known as Floquet engineering, we experimentally realize a long-ranged, anisotropic Heisenberg model with tunable interactions in a trapped ion quantum simulator. We demonstrate that the spectrum of the model contains not only single magnon excitations but also composite magnon bound states. For the long-range interactions with the experimentally realized power-law exponent, the group velocity of magnons is unbounded. Nonetheless, for sufficiently strong interactions we observe bound states of these unconventional magnons which possess a non-diverging group velocity. By measuring the configurational mutual information between two disjoint intervals, we demonstrate the implications of the bound state formation on the entanglement dynamics of the system. Our observations provide key insights into the peculiar role of composite excitations in the non-equilibrium dynamics of quantum many-body systems.
\end{abstract}

\maketitle

\section{Introduction}

Recent conceptual and technical progress has enabled the controlled study of quantum magnetism with quantum simulators and quantum processors. For example, the dynamics of magnons, which are itinerant flipped spin excitations that propagate through the system by the superexchange mechanism~\cite{Bethe1931,Auerbach_2012, Ganahl2012}, has been probed in quantum gas microscopes~\cite{Fukuhara_2013}.
Already in the early 1930s it has been realized
that in addition to isolated magnons, bound states of magnons exist as low-energy excitations of a ferromagnet~\cite{Bethe1931}. 
For short-range superexchange interactions, the remarkable light-cone dynamics of magnon bound states has been investigated~\cite{Fukuhara_2013_b} and their robustness has recently been demonstrated as well~\cite{Morvan2022}. Long-range spin interactions significantly modify this picture. When considering long-range spin flip-flop interactions, as realized by trapped ions in a strong transverse field~\cite{Jurcevic_2014, Richerme_2014, Monroe2021}, magnons can acquire an unbounded group velocity~\cite{Hauke_2013, jurcevic2015}. Hence, they are rather exotic quasiparticles. 

This raises the question of whether interactions between magnons, which enable the bound-state formation in the short-range model, can also stabilize bound states in a long-range spin chain. Recent theoretical work suggested that this is indeed the case~\cite{Macri_2021}. It remains an open challenge to experimentally realize and control models which allow for the observation of such bound states. In particular, scaling of quantum simulators to tens of trapped ions requires considerable technological developments. It is not obvious whether a model can be experimentally implemented with sufficient fidelity for large systems and long-times in the presence of noise which is inevitably present in any experiment. The anisotropic, long-range Heisenberg model can be implemented with trapped ions via Floquet engineering~\cite{Bukov_2015}. However, care has to be taken to design Floquet sequences that protect the quantum state against ambient noise and rotation errors.

\begin{figure*}[t!]
\centering
	\includegraphics[width=1.0\textwidth]{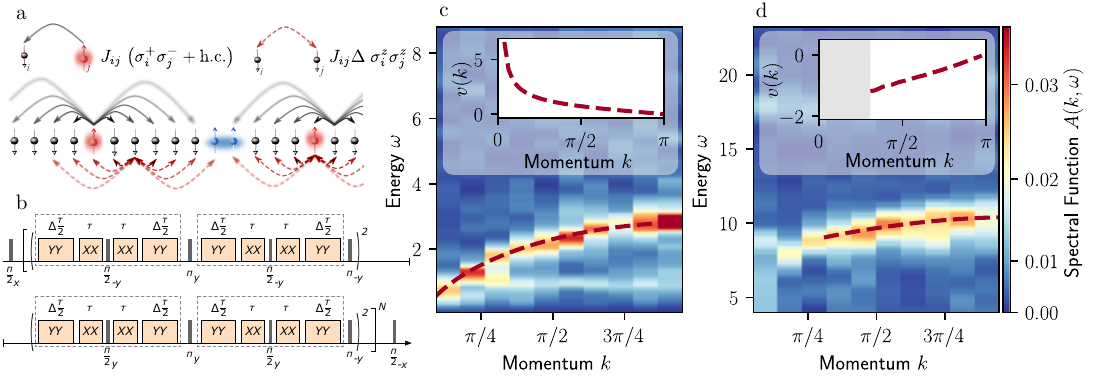}
	\caption{\textbf{Floquet engineering of the long-range, anisotropic Heisenberg model and measured dispersion relations.} (a)  The low-energy excitations of the ferromagnetic state in the long-range, anisotropic Heisenberg model consist of isolated magnons (red) and bound states thereof (blue). These excitations move and interact with each other by the long-range spin exchange couplings.
	(b) The Hamiltonian is experimentally realized from conventional long-range Ising interactions (illustrated by blocks of XX and YY) by Floquet engineering, which induces rotational symmetry in the XY plane of the spin and tunable ZZ interactions.
	(c) Plane-wave spectroscopy of the single magnon excitation energy indicates a cusp singularity at low momentum consistent with the theoretical prediction (red dashed line). Inset: The group  velocity $v(k)$ diverges in the long-wavelength limit.
    (d) Plane-wave spectroscopy of two-magnon states measured for strong interactions $\Delta = 3$ shows a well-defined resonance at large momenta. Measurements agree well with theoretical computations (red dashed line).   
	Inset: The velocity of the magnon bound state stays finite even when approaching the region of instability at small momenta, indicated by the gray area.}
	\label{fig_1}
\end{figure*}

In this work, we overcome these challenges and investigate bound-state dynamics in the long-range, anisotropic Heisenberg model realized in a linear string of ions, where two electronic states of the ion represent a spin-1/2. Spin-spin interactions can be engineered by illuminating the ions with a bichromatic laser beam.
This coupling gives rise to an effective long-range Ising interaction $H_\text{XX} = \sum_{i<j} J_{ij} \sigma_i^x \sigma_j^x $~\cite{Porras_2004} with spin-spin couplings that decay approximately algebraically with distance, $J_{ij} = J/|i-j|^\alpha$.
A key challenge is to enrich the set of available spin-spin interactions, starting from the long-range Ising interactions. By applying a strong transverse field, obtained by introducing either a third laser frequency component or by shifting the frequency of the bichromatic beam, spin-changing processes are penalized. This effectively realizes the long-range XY model describing flip-flop dynamics of spins \cite{Jurcevic_2014, Richerme_2014}. Enhancing the symmetry of the spin interactions from a discrete Ising to a continuous XY symmetry, enabled for example the spectroscopy of quasiparticle excitations~\cite{jurcevic2015}, the variational simulation of target Hamiltonians with related symmetries~\cite{Kokail:2019}, the study of hydrodynamic spin transport~\cite{Schuckert2022}, and the observation of spontaneous breaking of a continuous symmetry~\cite{Feng_2022}. However, in the long-range XY model direct interactions between magnons, which are necessary for the bound state formation~\cite{Macri_2021}, are absent. 
In order to realize these interactions between magnons, we resort to Floquet engineering~\cite{Bukov_2015}, and implement basis rotations such that the averaged Hamiltonian in the high frequency limit yields the long-range, anisotropic Heisenberg model
\be\label{eq_long_XXZ}
\hat{H}=\frac{1}{3}\sum_{i<j}J_{ij}(\hat{\sigma}_i^x\hat{\sigma}^x_j+\hat{\sigma}_i^y\hat{\sigma}^y_j+\Delta\hat{\sigma}_i^z\hat{\sigma}^z_j)\, ,
\ee
where $\hat{\sigma}^{x,y,z}$ are the Pauli matrices and $\Delta$ determines the Ising interactions between the magnons; see \figc{fig_1}{a}.
Several implementations of long-range Heisenberg models have been proposed; see for example Refs.~\cite{Arrazola_2016, Bermudez_2017, Birnkammer2022}.

\section{Experimental Realization}

In our experiments, long-range Heisenberg models are realized in a linear string of twenty ${}^{40}\mathrm{Ca}^+$ ions confined in a linear Paul trap with a low axial trapping frequency of 92~kHz and radial trapping frequencies of 2.926~MHz and 2.894 MHz, respectively. The ion crystal is laser-cooled by Doppler and polarization gradient cooling and its transverse modes are prepared close to their ground states by resolved sideband cooling. Qubits are encoded in the Zeeman levels $\left| \downarrow \right> = \left| S_{1/2}, m=+1/2 \right>$ and $\left| \uparrow \right> = \left| D_{5/2}, m=+5/2 \right>$ and manipulated with a narrow-linewidth laser at 729~nm. Single-qubit rotations are achieved by a steerable, tightly-focused laser beam inducing differential ac-Stark shifts between the qubit states of a single ion. In combination with a second laser beam resonantly coupling to all qubits with nearly the same strength, arbitrary single-qubit rotations can be achieved by sandwiching ac-Stark pulses between a pair of resonant $\pi/2$-pulses. Single-qubit addressing is used for preparing spatially structured initial states and for the measurement of spin-spin correlations \cite{kranzl2022}. 

The laser beam coupling to all ions is also used for engineering  effective Ising type spin-spin interactions. Entanglement between the qubits is generated by a bichromatic light field that off-resonantly couples the qubits to the collective motional modes from the direction perpendicular to the ion crystal. The wave vector of the light encloses a 45(5)$^\circ$ angle with each of the two radial principal axes of the harmonic trapping potential, leading to a coupling with all radial collective modes of motion. The frequencies of the radial modes cover a range of 137~kHz, the center-of-mass mode at 2.926~MHz having the highest frequency. The two frequency components of the laser beam are detuned by $\pm40$~kHz from the blue and the red sidebands of the highest-frequency mode of the ion crystal, giving rise to a long-range effective Ising coupling $H_\text{XX} = \sum_{i<j} J_{ij} \sigma_i^x \sigma_j^x$, that decays approximately as $J_{ij} = J/|i-j|^\alpha$ with a maximum coupling strength of $J = 369$~rad/s and a power-law exponent of $\alpha=1.4$; see Appendix~\ref{sec:app_powerlaw}  for a discussion on the choice of $\alpha$.

For the realization of the long-range, anisotropic Heisenberg model, we use a Floquet decomposition of the evolution operator, Fig.~\ref{fig_1}(b);  see Appendix \ref{sec:app_floquet} and also Ref.~\cite{Kranzl2022a}. The anisotropy $\Delta$ is controlled by setting the relative length of the Floquet substeps. To reduce the impact of dephasing noise, we implement the basis rotation between the Floquet substeps as dynamical decoupling pulses. Furthermore, the inhomogeneous coupling of the Gaussian laser beam, that induces collective spin rotations, causes rotation errors. These errors are  mitigated by alternating the direction of the basis rotations. The laser pulses of the individual Floquet steps are shaped to reduce spectral overlap with the radial motional modes. Related Floquet protocols which enhance the symmetries of the model by driving have for example recently been developed for Rydberg atoms~\cite{Geier2021, Scholl2022} and trapped ions~\cite{Kranzl2022a, Morong2022}. 

\begin{figure*}
	\includegraphics[width=1.0\linewidth]{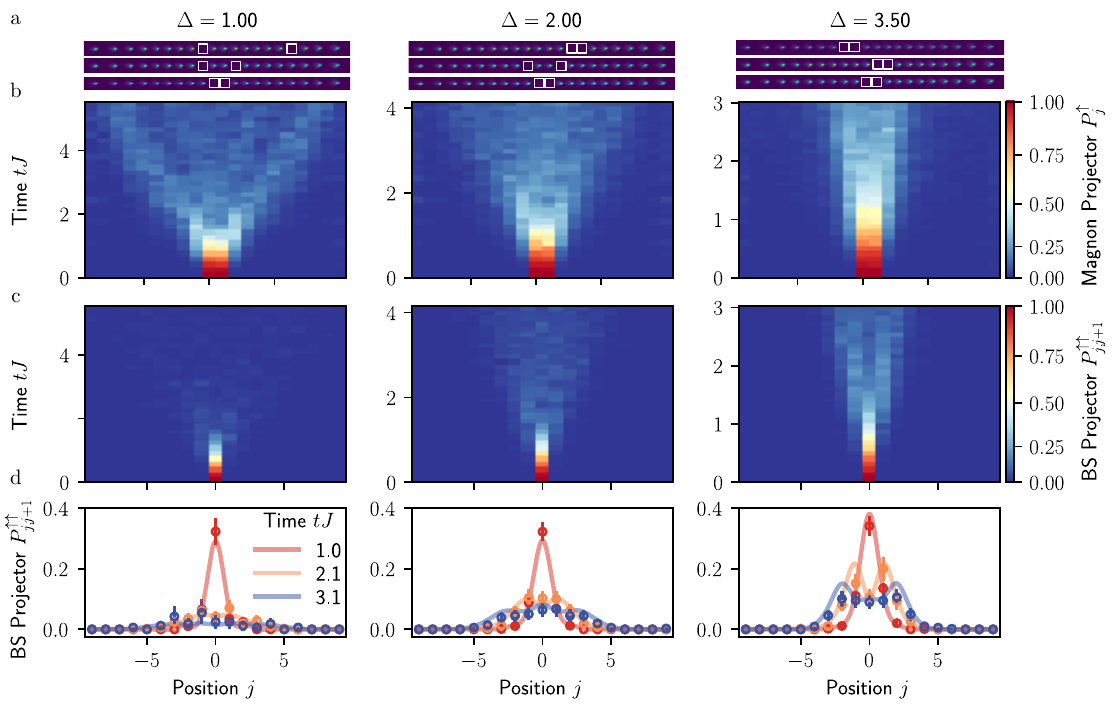}
	\caption{\textbf{Dynamics of magnon bound states.} We prepare two magnons in the center of the chain and study their evolution for interactions and maximum time $\Delta = 1.0$, $t_\text{max}J = 5.5$  (left), $\Delta = 2.0$, $t_\text{max}J  = 4.1$ (middle), and $\Delta = 3.5$, $t_\text{max}J  = 3.0$ (right), respectively.
	(a) Selection of three experimental snapshots measured at time $t = (0,t_{\mathrm{max}}/2,t_{\mathrm{max}})$ from bottom to top. For small interactions (left), magnons typically spread independently, whereas for intermediate and strong interactions (middle, right), two magnons are typically found nearby. (b) The one-magnon projector $P_j^\up$ spreads quickly for small interactions. It significantly slows down for strong interactions, for which it mainly captures the  bound state dynamics, indicated by a sharp light cone.
	(c) The two-magnon bound state (BS) projector $P_{j,j+1}^{\up\up}$ decays rapidly for weak interaction (left) but spreads with a light cone for larger interactions (middle, right) confirming the robust formation of a bound state of two nearby magnons.
   (d) The experimental measurements (symbols) for the two-magnon projector are in good agreement with the theoretical predictions (solid lines). Data is obtained from snapshots, which have been post-selected for two excitations. After post-processing we retain $\approx(22\%, 20\%, 25\%)$ of typically 1500 measured snapshots. Error bars are obtained from a jackknife analysis of the post-processed snapshots.
   }	
  \label{fig_2}
\end{figure*}

\section{Single- and two-magnon dispersion relations}

The dispersion of a single magnon excitation does not depend on the spin interaction $\Delta$, except for an overall constant offset. Using plane-wave spectroscopy introduced in Ref.~\cite{jurcevic2015} we characterize the dispersion $\varepsilon_1(k)$ of a single magnon (see \App{sec:spectroscopy} for details); \figc{fig_1}{c}. For all momenta in the Brillouin zone we measure a resonance, indicating the existence of a well-defined quasi-particle excitation. We find that the magnon velocity $v(k)=\partial_k \varepsilon_1(k)$ diverges in the long-wavelength limit, which becomes manifest by the cusp of the dispersion at zero momentum. The single-magnon dispersion can be readily computed analytically
$\varepsilon_1(k)= \frac{4J}{3}\sum_{\ell=1}^\infty \frac{\cos(k\ell)-\Delta}{\ell^\alpha}$ and is in good agreement with the measured data. This solution also shows that the maximal magnon velocity is unbounded for $1<\alpha\leq2$. The experimentally studied case of $\alpha=1.4$ is well within this range. As a consequence of the diverging group velocity, a local magnon excitation propagates arbitrarily fast.

\begin{figure}
	\includegraphics[width=1.0\columnwidth]{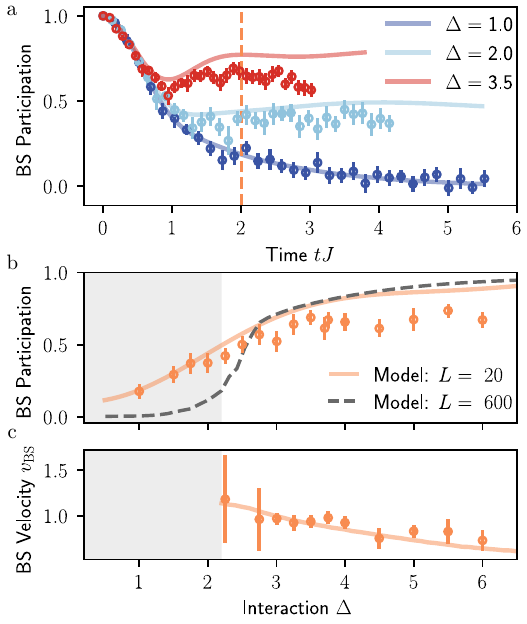}
	\caption{\textbf{Characterizing magnon bound states.} (a) For small interactions ($\Delta\!=\!1.0$) the bound state (BS) participation vanishes in time, resembling a random distribution of  magnons~(see \App{sec:BS participation}). By contrast, stable plateaus are reached at stronger interactions establishing the robust existence of the bound state.
	(b) The bound state participation at time $tJ=2.0$ [orange dashed line in (a)] unveils a crossover from a weak-interaction regime without bound states to a strong-interaction regime with magnon bound state at an interaction strength of $\Delta\approx2.2$.
	For comparison we show numerical data of the bound-state participation in larger systems evaluated at late times (gray dashed line).
    (c) For large interactions, we determine the velocity of the dominant bound state from the light cone spreading. Data is obtained from typically 1000 snapshots, which we post-select for two excitations. We retain $\approx 22\%$ of the snapshots at the latest times. Error bars are obtained from a jackknife analysis of the post-processed snapshots. 
	}
	\label{fig_3}
\end{figure}

Given the unconventional properties of the single-magnon excitation, in particular their unbounded velocity, it is \emph{a priori} unclear whether the model supports bound states of two magnons.
Here, we develop a technique for measuring the spectrum of two-magnon excitations by initializing the system approximately in a plane wave of two magnons situated next to each other, which can be achieved by perturbatively acting with long-range Ising interactions for a short period and tomographically reconstructing the two-site density matrix after an evolution with the long-range, anisotropic Heisenberg model (see \App{sec:two-mag spectroscopy}). We show in the appendix that the dominant bound state consists of two magnons situated next two each other, which motivates the preparation of the plane-wave initial state with two neighboring flipped spins. For extremely strong interactions, $\Delta \gg 1$, the long-range potential can host even further bound states, which are predominantly found at larger distances (see \App{sec:MultipleBS}). In general there are two competing effects: Increasing the interaction strength $\Delta$ favors the formation of bound states and lowering $\alpha$ which in turn increases the range of interactions tends to destabilize them (see \App{sec:app_powerlaw}). The results of this measurement of the dominant bound state is shown in \figc{fig_1}{d} for strong magnon interaction $\Delta = 3$. For low momenta, $k\lesssim \pi/4$, the spectral weight is distributed over a large energy window. By contrast, for high momenta, $k\gtrsim \pi/4$ a sharp resonance emerges, which is indicative of the formation of a bound state. 

We compare the measurement signal with the dispersion of the magnon bound state $\varepsilon_2(k)$, that we obtain from solving the two-body problem numerically, and find good agreement. The following features of the measurement should be emphasized: On the one hand, the bound state dispersion law does not stretch across the whole Brillouin zone, but becomes unstable upon approaching the long-wavelength limit. On the other hand, the group velocity of the composite two-magnon excitation remains always finite, in particular also when approaching the unstable regime at low momenta. As a consequence, the bound state of two magnons should possess a well-defined light cone. This is in stark contrast with the single magnon case.

\section{Dynamics of magnon bound states}

In order to establish the well-defined light cone of the magnon bound state, we prepare an initial state with two flipped spins situated next to each other in the center of the spin chain. Subsequently, we probe the non-equilibrium dynamics of this initial state under the evolution of Hamiltonian \eqw{eq_long_XXZ} for different values of the interaction $\Delta$. 
The snapshots measured at times $t = (0,t_{\mathrm{max}}/2,t_{\mathrm{max}})$ already exhibit striking signatures of the bound state formation; \figc{fig_2}{a}: At weak interactions $\Delta = 1.0$, the two magnons propagate freely and are typically measured at uncorrelated positions in space. By contrast, when increasing the interactions to $\Delta =2.0$ or even to $\Delta = 3.5$, we find them with high probability next to each other.

To quantitatively analyze the situation, we measure the single-magnon $P^{\uparrow}_j$ and two-magnon projectors $P^{\up\up}_{j,j+1}$ by averaging over a collection of typically $1500$ snapshots. The main contribution to the bound state results from two magnons located nearby, which is why we evaluate the two-magnon projector on adjacent sites in the spin chain (see \App{sec:BSphasespace}). As a consequence of the diverging group velocity, single-magnon excitations propagate faster than linear in time in the thermodynamic limit. For the accessible system size, this results in a broad distribution of the one-magnon projector for small interactions ($\Delta\!=\!1.0$).
The two-magnon projector, by contrast, quickly decays in time and does not carry substantial weight in the wave function, which indicates the absence of a magnon bound state; \fig{fig_2} (left column). 

When increasing the interactions to $\Delta = 2.0$ or even to $\Delta = 3.5$ a distinct behavior is observed; \fig{fig_2} (middle and right column, respectively). First, the one-magnon projector spreads slower than for $\Delta = 1.0$. This is incompatible with having a significant contribution of single unbound magnons in the wave function, as their dynamics are insensitive to the value of the interaction $\Delta$. As a consequence, the observed signal can be attributed to the formation of a magnon bound state. This is further confirmed by measuring the two-magnon projector, which displays a well-defined light cone with a velocity that is consistent with the one obtained from the single magnon projector. 
To further support that the effective time evolution of the Floquet protocol describes the long-range, anisotropic Heisenberg model, we compare time slices of the experimental data to numerical simulations and find excellent agreement, \figc{fig_2}{d}. 

\begin{figure*}
	\includegraphics[width=1.0\linewidth]{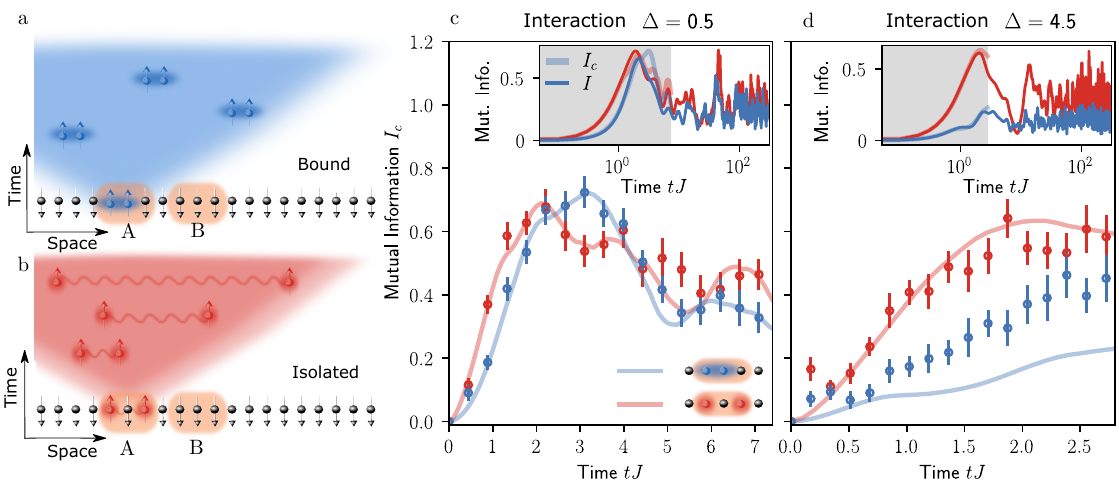}
	\caption{\textbf{Dynamics of mutual information.} We realize two distinct initial configurations, (a) one with two magnons next to each other and (b) one with two magnons separated in space. We characterize the spreading of configurational mutual information between segments A and B each consisting of three consecutive spins. 
	(c) For weak interactions ($\Delta\!=\!0.5$) the configurational mutual information $I_{c}$ yields similar results for both initial configurations.  (d) For strong interactions ($\Delta\!=\!4.5$), a significant difference in the growth of the mutual information is observed for the two initial configurations. Insets: Late-time dynamics and comparison of the measured configurational mutual information $I_{c}$ with the conventional mutual information $I$ (see text for details). Data is obtained from typically 2200 snapshots, which we post-select for two excitations. We retain $\approx 19\%$ of the snapshots at the latest times. Error bars are obtained from a jackknife analysis of the post-processed snapshots.
	}
	\label{fig_4}
\end{figure*}

We now analyze the magnon bound state formation for a large range of interaction strengths $\Delta$, by measuring total two-magnon projector $\sum_j P^{\up\up}_{j,j+1}$. This quantity not only captures the bound state but also two unbound magnons that accidentally occupy neighboring lattice sites. To correct for this effect we introduce the bound state participation, in which we normalize the two-magnon projector such that it discards contributions arising from a random placement of two magnons on neighboring sites. The random placement occurs with probability $2/L$, which is sizeable for the 20-ion chain considered here (see \App{sec:BS participation}). This bound-state participation then reaches a finite value at late times, provided a magnon bound state exists, and decays to zero otherwise.

In agreement with the light-cone measurements, we find a vanishing bound state participation for $\Delta\!=\!1.0$ at late times, and a saturation for larger values of the interaction $\Delta$; \figc{fig_3}{a}. The bound state contribution increases continuously with increasing interactions; \figc{fig_3}{b} and is in agreement with our numerical computations. The crossover  in the bound state participation becomes increasingly sharp, as we increase the system size.
In \App{sec:BSphasespace} we study the finite-size dependence and find a crossover in the spectrum of the model: bound states exist for $\Delta\gtrsim 2.2$. This is consistent with recently reported theoretical results~\cite{Macri_2021}.
We extract the light-cone velocity $v_{\mathrm{BS}}$ of the bound state from the propagation front of the two-magnon projector, which we identify by the half-maximum of the signal; \figc{fig_3}{c}.

\section{Dynamics of mutual information}

In systems with well-defined quasi-particles, the growth of entanglement following a quantum quench can be understood as follows: The initial state creates entangled quasi-particle excitations that propagate ballistically through the the system \cite{Calabrese_2005,Alba_2017}. In the long-range, anisotropic Heisenberg model two competing effects arise. On the one hand, single magnons spread arbitrarily fast, which could lead to entanglement growth that is faster than linear in time. On the other hand, composite bound states of magnons form, which strongly modify the excitation content of the state. In order to elucidate the consequences of the bound-state formation on the entanglement growth, we study the dynamics of two distinct initial states: one with two magnons situated next to each other [\figc{fig_4}{a}] that have a large overlap with the bound state provided the interaction is large enough, and one with two magnons that are separated by one lattice site [\figc{fig_4}{b}].  We then study the time evolution of this state under the long-range, anisotropic Heisenberg model and quantify the entanglement between two separated intervals $A$ and $B$. 

One way of quantifying the entanglement between two separated subregions that are embedded in a larger environment is achieved by the  mutual information $I=S_{A} + S_{B} - S_{A\cup B}$, where $S_A$ is the von Neumann entanglement entropy between the interval $A$ and the rest of the system $\bar A$. The entanglement entropy can be measured tomographically~\cite{Roos2004,Jurcevic_2014} and its R\'enyi variants by randomized measurements~\cite{Brydges2019} and beam splitter operations~\cite{Islam_2015}. However, proxies for the mutual information can already be obtained from the snapshots in one computational basis. Following Ref.~\cite{Lukin2019}, the basic idea is to observe that for systems which conserve the number of excitations $N$, the total density matrix has a block diagonal form $\hat{\rho}=\sum_{n=0}^N p(n) \hat{\rho}_A^{(n)}\otimes \hat{\rho}_{\bar{A}}^{(N-n)}$, where $p(n)$ is the probability of having $n$ magnons in the subsystem $A$, and $\hat{\rho}^{(n)}_A$ is the (normalized) reduced density matrix of the subsystem $A$ in the $n$-magnon sector. Then the total entanglement entropy $S_A$ can then be split into a number entropy $S_{N;A}$ and a configurational entropy $S_{C;A}$
\be\label{eq_entropy}
S_{A}=S_{N;A}+S_{C;A},
\ee
where $S_{N;A}=-\sum_{n=0}^N p(n)\log(p(n))$ and   $S_{C;A}=-\sum_{n=0}^N p(n)\text{Tr}(\hat{\rho}^{(n)}_A\log \hat{\rho}^{(n)}_A)$~\cite{Wiseman_2003,Goldstein_2008}. Measuring the configurational entropy exactly has an exponential cost. Hence, we approximate it for short times and small correlations between $A$ and $\bar{A}$ by the mutual configurational entropy~\cite{Lukin2019,Wolf_2008}
\be
S_{C;A}\simeq \sum_n p(n)\sum_{\{A_n\},\{\bar{A}_n\}}\big[p(A_n\otimes \bar{A}_n)-p(A_n)p(\bar{A}_n)\big]
\ee
where $\{A_n\}$ $(\{\bar{A}_n\})$ is the set of all the possible magnon configurations in the local basis in the subsystem $A$ $(\bar{A})$.
Using this approximation yields the proxy $I_{c}$ for the mutual information, which we measure in the experiment. 

For weak interactions, bound states cannot form. Hence, the initial state with two magnons next to each other and the one in which the magnons are separated by one site should give rise to comparable entanglement dynamics, as in both cases magnons propagate freely.  We find that both initial states give rise to comparable dynamics of the configurational mutual information $I_c$; \figc{fig_4}{c}, confirming the independent spreading of magnons. In the inset, we compare the theoretically computed configurational mutual information $I_c$ with the mutual information $I$ obtained from the von Neumann entanglement entropy and find good agreement between the two. 

For strong interactions the situation drastically changes. In that case, a robust magnon bound state forms, which reduces the growth of entanglement, as the bound state behaves effectively as a single particle. By contrast, when initializing the dynamics with two magnons separated by one lattice site, the magnons remain mainly unbound and move freely through the system, which increases the dynamically accessible number of configurations, hence leading to higher entanglement; \figc{fig_4}{d}. The insets of \figc{fig_4}{c,d} confirm that the observed behavior persists at late times: the mutual information $I$ saturates to a comparable value for weak interactions but is distinct for strong interactions, demonstrating the impact of magnon bound states on the entanglement dynamics. 

\section{Outlook}

By Floquet engineering, we have realized the long-range, anisotropic Heisenberg model in a linear crystal of 20 ions. The periodic drive enhances the discrete Ising interactions realized by sideband modulation to continuous Heisenberg-type interactions. This enabled us to study the interaction between magnons created on top of a ferromagnetic state. We experimentally found robust bound states for strong interactions.

The formation of bound states can have direct implications on the relaxation dynamics following a quantum quench in which a finite density of excitations is created. In our system bound states will be stable against few-body scattering processes for large interactions $\Delta$ due to  energy conservation. We expect that in this regime the relaxation will proceed in multiple stages: First, the system relaxes to an ensemble in which bound states act as well-defined quasiparticles. Second, bound states can decay due to higher order processes giving rise to full thermalization. Other ergodic systems featuring long-lived composite particles, such as mesons in confined spin chains~\cite{Kormos_2017,Lerose2020QL, Birnkammer2022b} or doublons protected by strong interactions in Hubbard models~\cite{Winkler2006, Strohmaier2010, Sensarma2010, Abanin_2017}, are expected to undergo a similar prethermal relaxation dynamics. 
This multi-stage dynamics should also manifest itself in the entanglement growth as indicated by our measurements on mutual information spreading. In future work, it will be interesting to investigate such multi-stage relaxation dynamics in systems which support long-lived bound states.

\begin{acknowledgments}
{\par\textit{Acknowledgements:}}
We acknowledge support from the Deutsche Forschungsgemeinschaft (DFG, German Research Foundation) under Germany’s Excellence Strategy--EXC--2111--390814868 and DFG grants No. KN1254/1-2, KN1254/2-1, and from the Austrian Science Fund through the SFB BeyondC (F7110). Furthermore, The project leading to this application has received funding from the European Research Council (ERC)
under the European Union’s Horizon 2020 research and innovation programme (grant agreements No.
741541 and No. 851161), from the Munich Quantum Valley, which is supported by the Bavarian state government with funds from the Hightech Agenda Bayern Plus, and from the Institut f\"ur Quanteninformation GmbH. 
{\par\textit{Data and Informations availability:}} Raw data, data analysis, and simulation codes are available on Zenodo~\cite{zenodo}.
\end{acknowledgments}

\textit{Author contributions}. The research was devised by MK and CFR. SB, AB, and MK developed the theoretical analysis. MJ, FK, RB, and CFR contributed to the experimental setup. FK and MJ performed the experiments. FK, SB, AB, MK and CFR analyzed the data and SB carried out numerical simulations. FK, SB, AB, MK, and CFR wrote the manuscript. All authors contributed to the discussion of the results and the manuscript. All correspondence should be addressed to MK (michael.knap@ph.tum.de) and CFR (christian.roos@uibk.ac.at).

\textit{Competing interests.} The authors declare no competing interests.

\appendix

\section{Floquet Hamiltonian realized with dynamical decoupling}
\label{sec:app_floquet}

\begin{figure}
	\includegraphics[width=1.0\columnwidth]{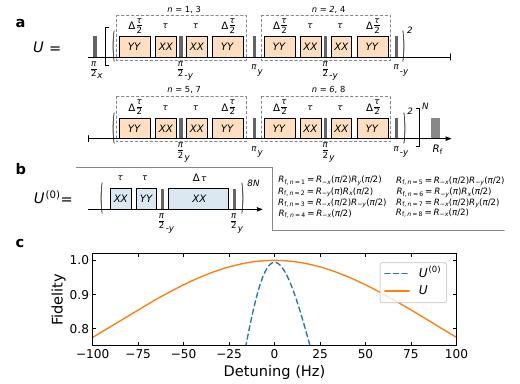}
	\caption{\textbf{Floquet protocol under dephasing errors}. 
	(a) Floquet Hamiltonian realized with dynamical decoupling. The series of Floquet steps (dashed boxes) repeats after $n=8$ steps. The final rotation $R_\mathrm{f}$ depends on the number of Trotter steps. Here, $R_\alpha(\theta)$ denotes a rotation around the axis $\alpha$ by an angle $\theta$.
	(b) Floquet Hamiltonian without dynamical decoupling.
	(c) We evaluate the robustness of the Floquet protocol under a dephasing error $\delta/2 \sum_i \sigma_z^{(i)}$, where $\delta$ denotes the detuning between laser and qubit. The numerical simulation is carried out for a system size of 10~qubits, magnon interaction $\Delta=3.5$ and a power-law spin-spin coupling $J_{i,j}=J/\left| i-j \right|^\alpha$ with $J=369$~rad/s and $\alpha=1.4$. The maximum evolution time is $tJ \approx 3.3$, divided into 32 Trotter steps.
	}
	\label{fig_6}
\end{figure}

The Floquet sequence uses dynamical decoupling to reduce the impact of dephasing errors caused by variations in the ambient magnetic field. Furthermore, global rotations are applied in an alternating fashion to reduce the impact of systematic over- and under-rotations, respectively, caused by inhomogeneous illumination by the global laser field. A numerical simulation of the Floquet sequence shows the robustness against constant detuning errors (Fig.~\ref{fig_6}). The dynamically decoupled sequence $U$ maintains a high fidelity over a much larger detuning range than the simple Floquet sequence $U^{(0)}$.

As a consequence of the dynamical decoupling pulses, the basis of the quantum state is rotated with each Floquet step. If the number $n$ of Floquet steps is not an integer multiple of eight (periodicity of Floquet sequence), the basis has to be rotated accordingly. The final rotation $R_\mathrm{f}$ is (operation on the right applied first)
$R_{\mathrm{f},n=1} = R_{-x}(\pi/2) R_{y}(\pi/2)$,
$R_{\mathrm{f},n=2} = R_{-y}(\pi) R_{x}(\pi/2)$,
$R_{\mathrm{f},n=3} = R_{-x}(\pi/2) R_{-y}(\pi/2)$,
$R_{\mathrm{f},n=4} = R_{-x}(\pi/2)$,
$R_{\mathrm{f},n=5} = R_{-x}(\pi/2) R_{-y}(\pi/2)$,
$R_{\mathrm{f},n=6} = R_{-y}(\pi) R_{x}(\pi/2)$,
$R_{\mathrm{f},n=7} = R_{-x}(\pi/2) R_{y}(\pi/2)$, and
$R_{\mathrm{f},n=8} = R_{-x}(\pi/2)$. Here, $R_\alpha(\theta)$ denotes a rotation around the axis $\alpha$ by an angle $\theta$.

For the experimental implementation of the Floquet sequence, an important source of error is the resonant coupling of the entangling light field to the radial motional modes of the ion crystal. As each Trotter substep is realized by a finite-length light pulse, the light's spectrum of the Floquet sequence is Fourier-broadened, with distinct peaks spanning over several tens of kilohertz. To reduce overlap with the radial motional modes, we use pulse shaping for each Trotter substep (Blackman window, rise and fall time~18 µs, respectively).

\begin{figure*}[t!]
	\includegraphics[width=1.0
	\linewidth]{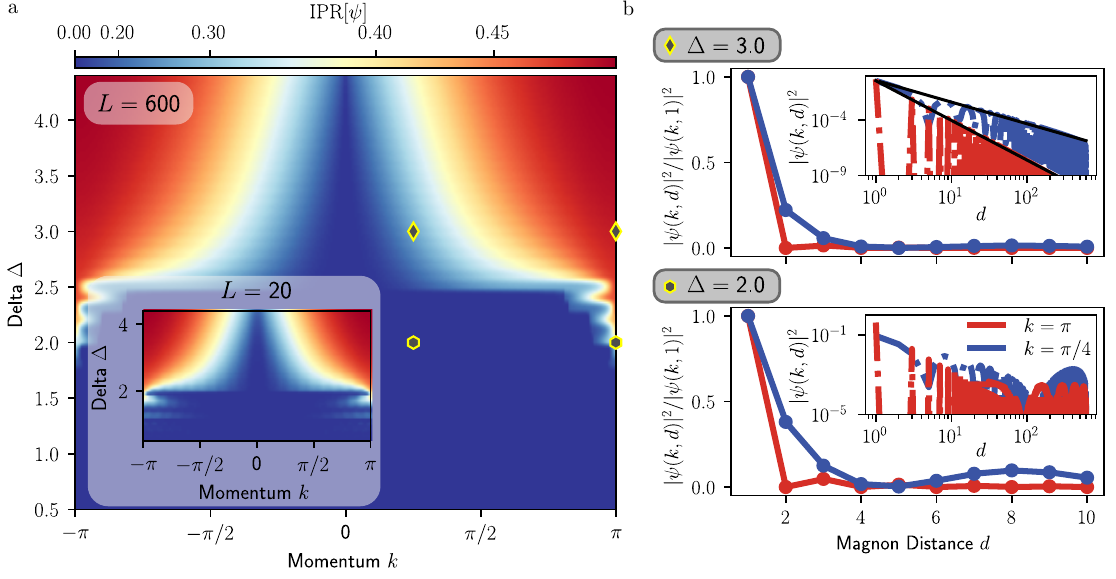}
	\caption{\textbf{Theoretically characterizing bound states}. (a) We show the inverse participation ratio (IPR) of the two-magnon eigenstates as a function of momentum $k$ and interaction $\Delta$ for comparatively large systems of $L=600$. A finite IPR indicates the existence of bound states, which arise in some part of the Brillouin zone for $\Delta\gtrsim 2.2$. The momentum range in which bound states exist increases with $\Delta$. Inset: IPR for the experimentally studied system size of $L=20$. (b) Examples of the eigenstates for large interaction $\Delta=3.0$ (upper panel) and intermediate interaction $\Delta=2.0$ (lower panel). When bound states exist (upper panel), the dominant contributions to the wave functions results from neighboring sites in the spin chain, as indicated by sharp peaks of $\vert\psi(k,d)\vert^2 / \vert\psi(k,1)\vert^2$.    
	}
	\label{fig_5}
\end{figure*}

\section{One-magnon spectroscopy}
\label{sec:spectroscopy}
In order to examine the dispersion law of single magnon excitations $\varepsilon_{1}(k)$ we use plane-wave spectroscopy. We start by creating a plane-wave superposition state of single magnon excitations with given momentum on top of the ferromagnetic  configuration $\ket{0}=\ket{\dw\dw...\dw\dw}$.
Plane-wave initial states can  be readily implemented using a sequence of modulated single qubit rotations mixing both locally accessible states of the computational spin basis. For our purpose such a rotation can be generated by individual single-ion rotations around the X-axis, 
\be
\ket{\psi_1}=e^{i\sum_j \gamma \mathcal{A}^{(k)}_j \hat{\sigma}_j^x}\ket{0} = \ket{0} + i \gamma \sum_j\mathcal{A}^{(k)}_j \hat{\sigma}_j^x\ket{0} + \mathcal{O}(\gamma^{2}).
\ee
Our goal is to excite plane waves of the form $\ket{k}=\frac{1}{\sqrt{L}}\sum_j e^{i kj} \hat{\sigma}_j^x\ket{0}$. For the open boundary conditions realized in the experiment the momentum is quantized as $k = \pi n/(L+1) $ for $n = 1, \ldots, L$. To this end, we choose the rotation angle to create a superposition of counter-propagating waves with $\mathcal{A}^{(k)}_j = \sqrt{\frac{2}{L}}\sin(kj)$.
In the measurements, we chose $\gamma=0.7$ and then postprocess for single magnon excitations. This implementation scheme allows us to create arbitrary linear combinations of single magnon states. For example, a superposition of two standing waves with different momenta $k$ and $q$ can be realized by fixing $\mathcal{A}^{(k)}_j=\sqrt{\frac{2}{L}}(\sin(kj)+\sin(q j))$, from which relative frequencies between the excitations at momentum $k$ and $q$ are obtained. This has the major advantage, that measurements are performed within a single magnon number sector, and hence the experiment operates in the subspace that is free of decoherence between different magnon number sectors.

Having implemented a suitable initial state, we continue by evolving it under the long-range, anisotropic Heisenberg Hamiltonian~\eqref{eq_long_XXZ}.
%\eq{eq_long_XXZ}. 
Because the dispersion law of a single magnon is expected to be independent of the chosen interaction $\Delta$, we consider the experimentally simplest case of vanishing interaction $\Delta=0$. Next, we perform spectroscopic measurements of the magnon projector $P_j^\uparrow=\frac{1}{2}(\hat{\sigma}_j^z+1)$ in the evolved state. The measured signal is
\begin{gather}
\bra{\psi_1 (t)}P_j^\uparrow\ket{\psi_1(t)}=\gamma^2 \; 2\Re\Big[e^{-it (\varepsilon_{1}(k)-\varepsilon_{1}(q))} \nonumber\\
\times \big(\bra{q}P_j^\uparrow\ket{k} 
+\bra{-q} P_j^\uparrow\ket{k}\big)\Big] + \mathcal{O}(\gamma^{3}),
\end{gather}
where non-oscillating contributions are dropped. 
We furthermore took advantage of the symmetry of the dispersion law under sign changes of the momentum. The energy difference $\varepsilon_{1}(k)-\varepsilon_{1}(q)$ is then obtained by a Fourier transform. The results of~\figc{fig_1}{c} 
correspond to a choice of the reference momentum $q=\pi/(L+1)$ as the standing wave of lowest frequency supported by the chain.

\section{Two-magnon spectroscopy}
\label{sec:two-mag spectroscopy}

A straightforward generalization of the one-magnon spectroscopy to two magnons requires the coherent manipulation of pairs of spins, which is difficult in our experimental setting. To solve this challenge we proceed by creating a coherent superposition of two-magnon states utilizing the short-time evolution of an Ising Hamiltonian $\sum_{j<j'} J_{jj'} \sigma_{j}^{x}\sigma_{j'}^{x}$ over a period $t J=0.19$. The coupling matrix decays as $J_{jj'}=J/\vert j'-j\vert^{\alpha}$. As a result we will predominantly obtain two-magnon states with both magnons situated at neighboring sites. Configurations with larger separation $d$ of the magnons will be suppressed. To realize a plane-wave initial state of two close-by magnons we have to furthermore tune the relative phase between the individual configurations. This can be achieved by applying a sequence of single qubit rotations with angles $\{\phi_{j}\}$ from the left-most to the right-most ion. As a result, we realize a two-magnon initial state  
\begin{multline}
\ket{\psi_2} = e^{-i \sum_j \frac{\phi_j}{2}\hat{\sigma}_j^z}e^{-i \gamma\sum_{j<j'} J_{jj'} \hat{\sigma}^x_j\hat{\sigma}^x_{j'}}\ket{0}\approx e^{i \sum_j \frac{\phi_j}{2}}\\ \times \Big(\ket{0}-i\gamma J_{12}\sum_j e^{-i(\phi_j+\phi_{j+1})}\ket{... \uparrow_{j}\uparrow_{j+1}...}\Big)+\mathcal{O}(\gamma^2)\,.
\end{multline}
By judiciously choosing the phase imprinting $\phi_j+\phi_{j+1}=j k$, we can target the desired plane wave. This state is then evolved with the long-range, anisotropic Heisenberg Hamiltonian~\eqref{eq_long_XXZ}.
%(\ref{eq_long_XXZ}). 
We obtain the energy of the two-magnon plane-wave configuration by measuring $\hat{\sigma}^-_{j}\hat{\sigma}^-_{j+1}$, where $\hat{\sigma}^{-}_{j}=(\hat{\sigma}^{x}_{j}-i\hat{\sigma}^{y}_{j})/2$. 
This observable connects configurations of two magnons located next to each other with the ferromagnetic state $\ket{0}$
\begin{gather}
\bra{\psi_{2}(t)}\hat{\sigma}^-_{j}\hat{\sigma}^-_{j+1}\ket{\psi_{2}(t)} = -i \gamma \Gamma J_{12} e^{-i jk} e^{-i \big(\varepsilon_{2}(k) -\varepsilon_{0}\big)t}\nonumber \\ + \;\mathcal{O}(\gamma^{2}).
\end{gather}
The measurement signal hence oscillates with the desired frequency of the two-magnon excitation. The contrast of the oscillation frequency carries also information about the probability amplitude $\Gamma$ of the bound state in the total many-body wave function.
Performing a Fourier transformation in time and averaging over sites 8 to 13 in the center of the chain to reduce contributions arising from the boundary, we obtain the dispersion law of the two-magnon bound state with respect to the ferromagnetic configuration $\varepsilon_{0} - \varepsilon_{2}(k)$, illustrated in~\figc{fig_1}{d}.

\begin{figure}[t!]
	\includegraphics[width=1.0
	\linewidth]{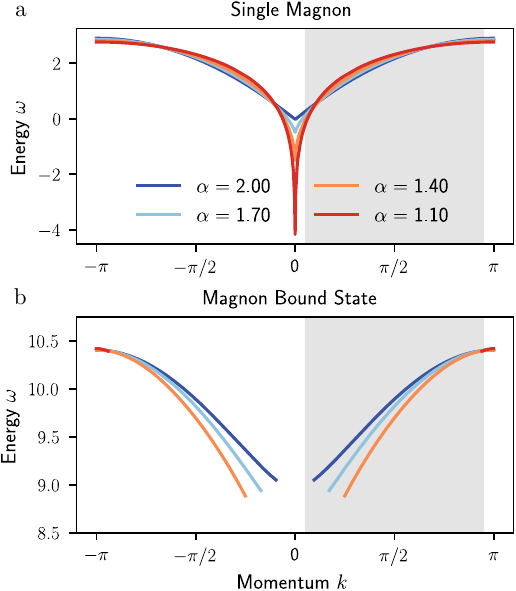}
	\caption{\textbf{Theoretical single magnon and bound state dispersions}. (a) Dispersion relations for a single magnon for values of $\alpha \in \{1.1, 1.4, 1.7, 2.0\}$. For $\alpha\leq 2.0$ dispersions feature a cusp in the limit of long wavelengths $k\to0$, archetypal to long-ranged models. Signatures for this characteristic are more pronounced for small $\alpha\to 1.0$, as reflected in higher curvature of the dispersion. (b) The bound state dispersion for magnon interaction $\Delta=3.5$ shows that the region in the Brillouin zone in which a stable bound state is formed is pushed toward higher momenta for decreasing $\alpha$. The spectroscopically accessible momentum range for the realized system size is shown as a gray shaded area. 
	}
	\label{fig_7}
\end{figure}

\section{Bound state participation}
\label{sec:BS participation}

The two-magnon projector $P^{\up\up}_{j,j+1}$ is sensitive to bound states, as they consist of correlated magnons at a short relative distance. When two unbound magnons are randomly placed on a chain of $L$ sites, a non-vanishing expectation value for the projector will arise statistically $\langle P^{\up\up}_{j,j+1}\rangle_{\text{random}}=2/L$. To account for this, we properly renormalize the bound state participation shown in~\fig{fig_3} 
as
\be
\text{BS participation} = \frac{1}{1-2/L}\Bigg( \sum_{j=1}^{L-1} \langle P_{j, j+1}^{\uparrow \uparrow}(t)\rangle - 2/L\Bigg).
\ee

\begin{figure*}
	\includegraphics[width=1.0
	\textwidth]{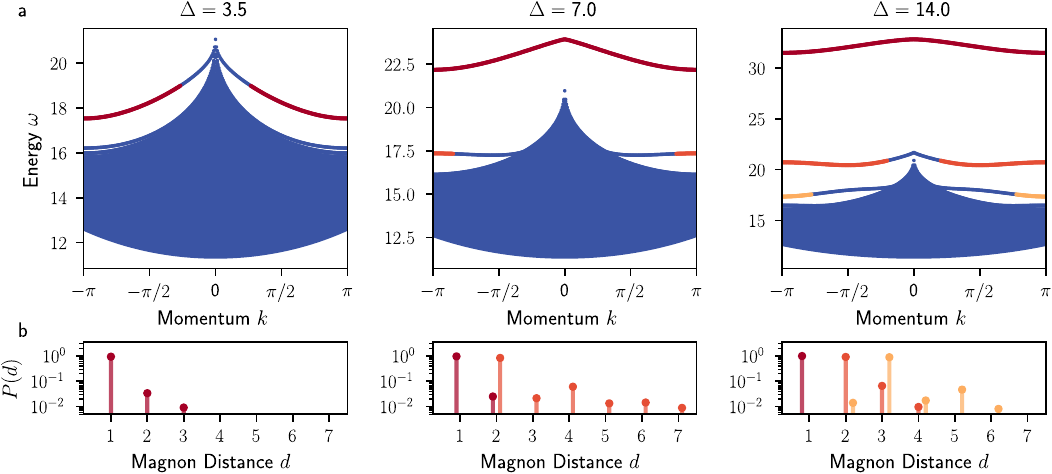}
	\caption{\textbf{Multiple bound states arising from long-ranged interactions}. (a) We compute the two magnon spectrum for a powerlaw exponent $\alpha=1.4$ and strong magnon interactions $\Delta\in\{3.5, 7.0, 14.0\}$ in a system of $L=500$ sites. For large $\Delta$ the powerlaw tail of interactions enables binding of magnons also at larger distances. In contrast to the short-ranged case one finds a hierarchy of different bound states in the spectrum. (b) Different bound states are thereby characterized by distinct magnon distances. This becomes apparent from computing the probability distribution $P(d)$ for finding the different types of bound states with magnon distance $d$. For all investigated values of $\Delta$ the individual distributions $P(d)$ are strongly peaked around a single distance $d$ (note the logarithmic scale). }
	\label{fig_10}
\end{figure*}

\section{Characterization of two-magnon bound states}
\label{sec:BSphasespace}

The results presented in the main text indicated a non-trivial dependence of bound states on the interaction $\Delta$ and momentum $k$.
We now provide a theoretical analysis of the eigenstates of the long-range, anisotropic Heisenberg model in the two-magnon sector.
A convenient strategy to detect localized wavefunctions is to estimate their spatial support using the inverse participation ratio (IPR). A finite IPR in the thermodynamic limit indicates a bound state.
Let $\psi(k,d)$ be a normalized eigenstate in the two-magnon sector with momentum $k$, and relative distance between the flipped spins $d$. Then the IPR is defined as
$\mathrm{IPR}[\psi(k)]=\sum_{d=1}^{L-1} \vert \psi(k, d) \vert^{4}$.

We obtain the wave functions numerically by exact diagonalization (ED) of the studied spin Hamiltonian $H$. We explicitly implement the magnon number conservation of the model. This allows us to treat sectors of the Hilbert space with different magnon numbers separately. The size of the sector capturing $m$ magnons is thereby given by the combinatorial factor $\binom{L}{m}$, which scales as  $\mathcal{O}(L)$ for one magnon  and as $\mathcal{O}(L^2)$ for two magnons. This enables us to perform numerically exact simulations for comparatively large systems up to $L=600$ sites.   

In \figc{fig_5}{a} we show results of the IPR for the state with highest energy, representing the probable candidate for a bound state in the case of repulsive magnon interactions ($\Delta>0$). For large systems ($L=600$) we find well-separated regions with finite IPR for $\Delta\gtrsim 2.2$. When increasing the interaction strength, the regime in which the bound states exist starts to extend over a larger momentum range but it always vanishes at the peculiar point $k=0$, i.e., at the point where the single-magnon velocity diverges.
Simulations carried out with the experimentally realized system size of $L=20$ yield qualitatively similar results, but a non-negligible contribution persists for smaller interaction $\Delta$; see \figc{fig_5}{a} (inset). 

We study the decay of the bound state wave function with relative distance between the magnons in \figc{fig_5}{b}. Within the region featuring stable bound states ($\Delta=3$, upper panel) we find that the bound state decays as a power law with distance in space (inset). The bound state therefore mainly occupies nearest neighbor sites, which motivates the choice of the two-site projector of magnons on adjacent sites, which we have used in the main text.
For smaller interactions ($\Delta=2$, lower panel) strong tails emerge in the wave function that do not decay to zero. Such wave functions give rise to a vanishing IPR and represent an extended state.

\section{Role of the powerlaw exponent}
\label{sec:app_powerlaw}

Our experimental results presented in the main text are obtained for the powerlaw exponent $\alpha=1.4$. In the following we will comment on why this value of $\alpha$ is favourable for our studies. Our goal is to elucidate the long-range character of our model. The single-magnon dispersion diverges at long wavelenths for powerlaw exponents of $1 < \alpha \leq 2$, which is archetypal for excitations in long-ranged systems. The theoretically expected dispersion relations of a single magnon as well as the two magnon bound state for this parameter regime are shown in \fig{fig_7} for a couple of $\alpha$. 
On the one hand, the long-ranged characteristics of the interactions become more pronounced when considering small values of $\alpha\to 1.0$, see \figc{fig_7}{a}. This limit has, however, profound consequences for the magnon bound state. When approaching $\alpha\to 1.0$ the regime in the Brillouin zone in which the bound state exists diminishes to a small momentum window around $k=\pi$; \figc{fig_7}{b}. This makes it much harder to detect and characterize the bound state experimentally.

On the other hand, considering the opposite limit $\alpha\to 2.0$ the bound state covers almost the whole Brillouin zone; \figc{fig_7}{b}. This regime is, however, challenging to access experimentally. This is because the larger the exponent, the weaker the absolute coupling strength $J$ at a given available laser power. Given the maximum coherence time of the experiment, this reduces the maximal effective time scales measured in units of $1/J$ that we can reach. The presented value of $\alpha = 1.4$ allows us to probe systems of reasonable sizes up to comparable long times. However, the same phenomenology would apply for all values of $\alpha$ in the range $1 < \alpha \le 2$.

\section{Multiple bound states arising from long-ranged interactions}
\label{sec:MultipleBS}

So far our results indicated that long-range interactions between magnons allow to stabilize a single bound state; a result qualitatively similar to the model only taking into account short-ranged interactions. In this section we show that the long-range interactions can stabilize more than one bound state. Increasing the value of interactions $\Delta$, which stabilizes our bound states, we identify two qualitative changes in the two magnon spectrum; \fig{fig_10}. First, we find additional bound state bands splitting off the continuum giving rise to a hierarchy of bound states in the strong coupling limit, see \figc{fig_10}{a}.
Secondly, we observe that the propability distribution functions $P(d)= \sum_{k}\vert\psi(k,d)\vert^2$, depicted in \figc{fig_10}{b}, are strongly peaked
at certain distances. This indicates that each bound state is located at a specific distance.
The tightest bound state, which is probed in the experiment, in particular, is situated on nearest neighbors. This further motivates the choice for the bound state projector $P^{\up\up}_{j,j+1}$ used to detect bound states in the measurements, as the dominant weight for the experimentally studied bound state wave function arises from neighboring magnons.

%\bibliography{biblio}

%

\end{document}